\documentclass[prd, a4paper, amsfonts, amssymb, amsmath, reprint, showkeys, nofootinbib, twoside]{revtex4-1}

\usepackage{slashed}
\usepackage[pdftex, pdftitle={Article}, pdfauthor={Author}]{hyperref} % For hyperlinks in the PDF
\newcommand{\chib}{\overline{\chi}}
\newcommand{\KD}{{K\"{a}hler-Dirac\;}}
\newcommand{\Psib}{\overline{\Psi}}
\newcommand{\Phib}{\overline{\Phi}}
\newcommand{\phib}{\overline{\phi}}
\newcommand{\psib}{\overline{\psi}}

\newcommand {\tr}{{\rm tr\,}}
\begin{document}
\title{'t Hooft anomalies for staggered fermions}

\author{Simon Catterall}
\email{smcatter@syr.edu}
\affiliation{Department of Physics, Syracuse University, Syracuse, NY 13244, USA.}

\begin{abstract} We show that the phase structure of
certain staggered fermion theories can be understood on the basis of
exact anomalies. These anomalies arise when staggered fermions are coupled
to gravity
which can be accomplished by replacing them
by discrete \KD fermions. We first show the existence of a perturbative
anomaly in even dimensions which breaks an exact $U(1)$ symmetry of the massless
theory down to $Z_4$. If we attempt to gauge this $Z_4$ symmetry we find a 't Hooft anomaly
which can only be cancelled for multiples of two \KD fields. This result is consistent with
the cancellation of a further mixed non-perturbative 't Hooft
anomaly between the global $Z_4$ and a reflection symmetry.
In four dimensional flat space, theories of two staggered fields yield eight Dirac or
sixteen Majorana fermions in the continuum limit and this critical number of fermions
agrees with results in condensed matter theory literature
on the fermion content required to gap boundary
fermions in $4+1$ dimensional topological superconductors. It is also consistent
with constraints stemming from the cancellation of 
spin-$Z_4$ anomalies of Weyl fermions. Indeed, cancellation of 't Hooft anomalies
is a necessary requirement for symmetric mass generation and this result
gives a theoretical explanation of recent numerical work
on the phase diagram of interacting staggered fermions. As an application of these ideas
we construct a lattice model whose low energy continuum limit is conjectured
to yield the Pati-Salam
GUT theory.
\end{abstract}

\date{\today} 
\maketitle

\section{Introduction}
Staggered fermions are a well known lattice discretization of 
relativistic fermions. Their use dates
from the earliest days of lattice gauge theory \cite{PhysRevD.11.395} and they
have since
become one of the most popular realizations of lattice fermions  employed in large scale
high precision simulations of QCD - see \cite{RevModPhys.82.1349} and references therein. Given this
history one might imagine that there was little new to
say concerning theoretical aspects of staggered
fermions. In this paper we argue that this is not the case; staggered
fermions are subject to certain gravitational
't Hooft anomalies which can be captured exactly in the lattice theory and which
play a role in determining the infrared dynamics in these theories.

Our work
was driven, in part, by recent observations of new phases of interacting staggered fermions in
which the fermions acquire mass without breaking any exact lattice symmetries
\cite{Ayyar:2014eua,Ayyar:2015lrd,Catterall:2015zua,Ayyar:2016lxq,Ayyar:2017qii,Butt:2018nkn}.
In this paper we will argue that the existence of such massive symmetric
phases requires the cancellation of certain 't Hooft anomalies.
To see these anomalies one needs to generalize
staggered fermions to curved space. Fortunately this is possible by exploiting
their known connection to \KD fermions. In spite of the need for curved
space to expose these anomalies we should stress that 
they have direct consequences for the possible
phases of staggered fermions in flat space. 

The plan of the paper is as follows. In the next section we present a lightning
summary of the essential features of 
staggered fermions. This is followed by a review of \KD fermions
including their connection to staggered fermions and their discretization on
random triangulations which are necessary to describe curved space. We then show
how discrete \KD fermions in even dimensions
suffer from an anomaly that breaks an exact global $U(1)$ symmetry
to $Z_4$. This global symmetry is a generalization of the usual $U(1)_\epsilon$ of staggered
fermions to a general random triangulation.

We then demonstrate how to gauge the \KD action under this
$Z_4$ symmetry and show that in general there is an obstruction
to this process indicating the presence of a
't Hooft anomaly. We then show that this anomaly can be cancelled
and the partition function rendered gauge
invariant if the theory contains
multiples of two \KD fields. Finally we use a spectral flow argument
to expose the presence of a further non-perturbative 't Hooft
anomaly in the system corresponding to a mixed anomaly between the $Z_4$ symmetry and
a reflection symmetry.  It can be seen by examining the propagation of
\KD fermions on non-orientable triangulations. 
We show that this anomaly can also be cancelled if the number of
\KD fields is zero mod 2.
Finally we use these results to construct a staggered lattice model
whose low energy limit we argue targets the Pati-Salam GUT theory.

\section{Review of staggered fermions}
In this section we briefly review the aspects of staggered fermions which
we will need for our later discussion of \KD fermions.
The free staggered fermion action takes the form
\begin{equation}
    S=\sum_{x,\mu}\eta_\mu(x)\chib(x)\Delta_\mu\chi(x)+\sum_x m\chib(x)\chi(x)
\end{equation}
where $\Delta_\mu$ is a symmetric difference operator defined by 
\begin{equation}\Delta_\mu f(x)=\frac{1}{2}\left(f(x+\mu)-f(x-\mu)\right)\end{equation}
and $\eta_\mu(x)=\left(-1\right)^{\sum_{i=1}^{\mu-1} x_i}$ are the well known staggered fermion phases which depend on the coordinates of sites 
in a hypercubic lattice. The staggered fields $\chi(x)$ and $\chib(x)$ are single component Grassmann fields. In addition to the simple $U(1)$ phase invariance corresponding
to $\chi\to e^{i\beta}\chi$ and $\chib\to \chib e^{-i\beta}$ the staggered
action at $m=0$ is also invariant under
an additional $U(1)$ symmetry which transforms the fields according to:
\begin{align}
    \chi(x)&\to e^{i\alpha\epsilon(x)}\chi(x)\nonumber\\
    \chib(x)&\to \chib(x)e^{i\alpha\epsilon(x)}
\end{align}
where $\epsilon(x)=\left(-1\right)^{\sum_{i=1}^d}$ is the site parity. Indeed in this limit
the action decomposes into two independent pieces
\begin{equation}
    S=\sum_{x,\mu}\eta_\mu(x)\chib_-(x)\Delta_\mu\chi_+(x) + \eta_\mu(x)\chib_+(x)\Delta_\mu\chi_-(x)
\end{equation}
where
\begin{equation}
\chi_\pm(x)=\frac{1}{2}\left(1\pm\epsilon(x)\right)\chi(x)
\end{equation}
If we retain just one of these pieces - say the first - we 
obtain a {\it reduced} staggered action in which each lattice site contains
a single reduced fermion field - $\chi_+$ on even lattice sites and $\chib_-$ for odd parity
sites. Renaming $\chib_-\to \chi_-$ we can then write the reduced action in the simple form
\begin{equation}
    S_{\rm red}=\sum_{x,\mu}\eta_\mu(x)\chi(x)\Delta_\mu\chi(x)
\end{equation}
 
The most common derivation of this action arises by 
spin diagonalizing the continuum Dirac action 
but it is better for our purposes to think of it as arising from
discretization of the continuum action for a \KD field. In the next section
we will review \KD fermions and their connection to staggered fermions. 

\section{{\KD} fermions on and off the lattice}

The Laplace-de Rham or \KD operator $d-d^\dagger$, where $d$ is the
exterior derivative, is a natural square root
of the Laplacian and can be used to write down an equation for fermions
which is an alternative to the Dirac equation \cite{kahler}:
\begin{equation}
    (d-d^\dagger+m)\Phi=0
    \label{kd}
\end{equation}
The \KD field $\Phi=(\phi,\phi_\mu,\ldots \phi_{\mu_1\ldots \mu_D})$ is a collection of antisymmetric tensor (p-form) fields defined on
a general $D$-dimensional Riemannian manifold. In flat space one can use these
forms to build a $2^{D/2}\times 2^{D/2}$ matrix $\Psi$ using elements of
the Clifford algebra:
\begin{equation}
    \Psi=\sum_{n=0}^D \phi_{\mu_1\ldots \mu_n}\gamma^{\mu_1}\cdots\gamma^{\mu_D}
\end{equation}
It is then a straightforward exercise to show that this matrix field satisfies
the usual Dirac equation and describes $2^{D/2}$ degenerate Dirac spinors
corresponding to the columns of $\Psi$ \cite{Banks:1982iq}.

The fact
that \KD fermions are equivalent to multiplets of Dirac fermions in flat space
ceases to be true in curved space. One way to see this is to recognize that the \KD equation eqn.~\ref{kd} is well defined for any smooth manifold - not just those supporting
a spin structure. In fact the global
properties of \KD fermions are very different from Dirac -- for example
the zero modes of the \KD equation on a compact space are
simply the harmonic forms. This means that there are zero
modes on spaces with positive curvature such as the sphere which
is not true of Dirac fermions. 

The \KD equation possesses another key
advantage over the Dirac equation; it can be
discretized {\it without} introducing spurious fermion doubler modes.~\footnote{Of course the resulting lattice theory leads to $2^{D/2}$ Dirac
spinors in the continuum limit but this is also a feature of the continuum theory - discretization does not add to this existing degeneracy.}

We assume that
any curved space can be approximated by a suitable
oriented triangulation and 
discretization proceeds by mapping continuum p-forms to fields defined on p-simplices in the
triangulation (sometimes called p-cochains in the literature).  Each $p$-simplex is specified by a list of $(p+1)$ vertex labels $\left[a_0,\ldots a_p\right]$ and the p-simplex field  $\Phi^{(p)}$ is given by the formal sum
\begin{equation}
   \Phi^{(p)}= \sum_{\rm p-simplices} \left[a_0,\ldots, a_p\right] \phi^{(p)}_{\left[a_0,\ldots, a_p\right]}
\end{equation}
Discrete analogs of $d$ and its adjoint $d^\dagger$ exist - the
co-boundary $\bar\delta$ and boundary 
$\delta$ operators  \cite{Rabin:1981qj,Becher:1982ud,Catterall:2018lkj,Catterall:2018dns} with
the action of $\delta$ on a p-simplex field being given by
\begin{equation}
    \delta_p {\Phi}^{(p)}=\sum_p\sum_{k=0}^{p} \left(-1\right)^k\left[a_0,\ldots\hat{a_k},\ldots ,a_p\right] 
    \phi^{(p-1)}_{\left[a_0,\ldots\hat{a_k},\ldots ,a_p\right]}\label{boundaryop}
\end{equation}
where $\hat{a_k}$ denotes the vertex that is removed to get the $k^{\rm th}$ boundary $(p-1)$-simplex. Using the abbreviated notation
\begin{equation}
C_p\equiv {\left[a_0,\ldots, a_p\right]}
\end{equation} we can write this as
\begin{equation}
    \delta_p \phi(C_p)=\sum_{C_{p-1}} I(C_p,C_{p-1}) \phi(C_{p-1})\label{incident}
\end{equation}
where $I(C_p,C_{p-1})$ is a $N_p\times N_{p-1}$ incidence matrix whose matrix elements are $+1$ if $C_{p-1}$ is
contained in the boundary of $C_p$ with the correct orientation, $-1$ if it occurs with
opposite orientation and zero otherwise.
Similarly
\begin{equation}
    \bar\delta_p \phi(C_p)=\sum_{C_{p+1}} I(C_{p+1},C_p)^\dagger \phi(C_{p+1})
\end{equation}
The Laplacian operator which maps $p$-simplex fields to $p$-simplex fields is then
\begin{equation}
I(C_{p},C_{p-1}) I(C_p,C_{p-1})^T+I(C_{p+1},C_{p})^T I(C_{p+1},C_{p})
\end{equation}
Zero modes of the \KD operator $\delta-\bar\delta$
are simultaneously (discrete) harmonic forms.

Finally an oriented triangulation is one in which each D-simplex is equipped
with a additional $Z_2$ element $\tau(C_D)$ which represents the orientation inside the simplex.
One can think of it as classifying whether a given vertex ordering is
an odd or even permutation of some fixed vertex ordering such as $a_0<a_1<\ldots a_D$.
The boundary operator is modified to $\delta_D\to \tau\delta_D$.
For an orientable triangulation $\tau(C_D)$
can be chosen in such a way that each face is held with opposite orientation
in the two D-simplices that share it. 

The discrete \KD equation is simply
\begin{equation}
    (\delta-\bar\delta+m)\Psi=0
\end{equation}
where $\delta=\sum_p \delta_{p}$.

Both the continuum and discrete {\KD} operators anticommute with an operator $\Gamma$
which acts on a given p-form field by multiplying
it by $\left(-1\right)^p$. This implies that
the associated {\it massless} \KD action is invariant under a $U(1)$ symmetry which acts as
\begin{align}
    \Phi&\to e^{i\alpha\Gamma}\Phi\nonumber\\
    \Phib&\to \Phib e^{i\alpha\Gamma}
\label{u1}
\end{align}
In the continuum its action on the matrix fermion $\Psi$ corresponds to
$\gamma_5\Psi\gamma_5$ and hence in flat space it is sometimes
called a ``twisted" chiral symmetry.
Using $\Gamma$ one can build projectors and reduced \KD fields $\Phi_\pm$
in complete analogy to our earlier discussion of reduced staggered fields. The
reduced \KD action for four reduced fields takes the form
\begin{equation}
    S_{\rm RKD}=\int \Phib_+\,K\,\Phi_-\quad{\rm with}\; K=\delta-\bar\delta
\end{equation}
We can rewrite this in a useful form by introducing the reduced field $\Psi^T=\left(\Phib_+\; \Psi_-^T\right)$ as
\begin{equation}
    S_{\rm RKD}=\int \Psi^T {\cal K}\Psi\quad {\rm where}\;{\cal K}=\left(\begin{array}{cc}0&K\\-K^T&0\end{array}\right)
\end{equation}

In flat space one can map the discrete p-form fields into 
fields defined on the p-cells of a regular hypercubic lattice rather than a triangulation. 
By introducing a new lattice with half the
lattice spacing one can replace these p-cell fields with single component fields on this
finer lattice according to the prescription
\begin{equation}
    \chi({\bf x}+\hat{\mu}_1+\hat{\mu}_2+\ldots+\hat{\mu}_p)=\phi_{\left[\mu_1\ldots \mu_p\right]}(x)
\end{equation}
From these we can form a discrete \KD matrix field using
\begin{align}
    \Psi(x)&=\sum \chi({\bf x}+\hat{\mu}_1+\ldots+\hat{\mu}_p)\gamma^{\mu_1}\cdots\gamma^{\mu_D}\nonumber\\
    &=\sum_b \chi(x+b)\gamma^{x+b}
\end{align}
where $b$ a vector whose components are either zero or one with the sum extending over
the unit hypercube of the lattice associated with
lattice site $x$. We now
plug this into
the continuum matrix action replacing integrals by lattice sums and the derivative
operator by a symmetric difference operator to obtain
\begin{align}
    S&=\sum_x \sum_{b,b^\prime}\chib(x+b)\chi(x+b^\prime+\mu){\rm Tr}\left[(\gamma^{x+b})^\dagger\gamma_\mu\gamma^{x+b^\prime+\mu}\right]\nonumber\\
    &-(\mu\to -\mu)
\end{align}
Evaluating the trace we find a factor of
$\eta_\mu(x+b)\delta_{b,b^\prime}$ and the action collapses to
the usual one for staggered fermions. It can be seen
that the staggered fermion phases $\eta_\mu(x)$ just reflect the
antisymmetry of the forms while
the operator $\Gamma$ becomes just the usual site parity $\epsilon(x)$. 
Notice that
this \KD description makes clear how to assemble the staggered fermion fields
to rebuild the continuum Dirac spinors -- they are once again given by columns of the matrix
field $\Psi$ in the given basis of Dirac matrices. At this point it should be clear
that the staggered fermion action is merely a form of the
discrete \KD action specialized
to a flat hypercubic lattice.

\section{\label{Z4} A perturbative anomaly}

The existence of $\Gamma$ ensures that the spectrum of the massless
\KD operator on a general random
triangulation pairs a state with eigenvalue $\lambda$ to another with
eigenvalue $-\lambda$ and ensures that zero modes of the \KD operator are 
eigenstates of $\Gamma$. Indeed, the \KD operator on a general triangulation
obeys an index theorem
\begin{equation}
    n_+-n_-=\chi
\end{equation}
where $n_\pm$ denotes the numbers of zero modes with $\Gamma=\pm 1$ and
$\chi$ is the Euler characteristic of the space. 

This index theorem can be related to an anomaly for the $U(1)$ symmetry given in eqn.~\ref{u1} by examining the
variation of the lattice \KD fermion measure under the
symmetry\cite{Catterall:2018lkj}. The latter is given by
\begin{equation}
    D\Phi\,D\Phib=\prod_{p=0}^D \prod_{i=1}^{N_p} d\phi_p(i)\,d\phib_p(i)
\end{equation}
Under $e^{i\alpha\Gamma}$ this transforms as
\begin{equation}
    D\Phi\,D\Phib \to e^{2i\alpha\left(N_0-N_1+\cdots +N_D\right)}D\Phi\,D\Phib=e^{2i\alpha\chi}D\Phi\,D\Phib
\end{equation}
where $\chi$ is the Euler character of the space.
Compactifying $R^D$ to $S^D$ where $\chi=2$ one discovers that this gravitational
anomaly breaks $U(1)\to Z_4$. Notice that the appearance of
an anomaly in this lattice theory requires only the existence of exact
zero modes carrying well defined charge under
the generator of the symmetry - it does {\it not}
require an infinite numbers of degrees of freedom - the anomaly can be computed
precisely on the coarsest triangulation of a manifold with a fixed topology. Ultimately
this results from the fact that the number and identity of zero modes of
the \KD operator are given by the ranks of the homology groups of the space which are
determined by its triangulation \cite{Rabin:1981qj}. Notice that
this remaining $Z_4$ symmetry is sufficient to
prohibit fermion bilinears from arising as quantum corrections to
the effective action but allows for four fermion terms.

One can think of this anomaly as a form of mixed 't Hooft anomaly corresponding to the
breaking of the global
$U(1)$ symmetry of staggered fermions in a non-trivial background gauge field corresponding to curved
space. 

The existence of this anomaly has important consequences for theories of {\it reduced} \KD fermions coupled
to a $U(1)$ gauge field and propagating on a curved background. The anomaly in this
case now reflects a violation of
gauge invariance and ruins the consistency of the theory.  This is analogous to the fact that the usual ABJ anomaly of Dirac fermions implies that it is
not possible to couple a single Weyl field to a $U(1)$ gauge field.
For a set of
reduced fields it can cancelled if the sum of their $\Gamma$ charges
vanishes. 

Alternatively the anomaly can be cancelled if the theory lives on the boundary of a higher dimensional
space where the bulk theory contains a topological Chern-Simons term whose anomalous gauge variation cancels that of
the boundary fermions. Details of such an anomaly inflow mechanism for the case of a three dimensional
bulk with two dimensional massless \KD boundary fermions were given in \cite{Catterall:2022ukg}. This construction
generalizes straightforwardly to yield a $(4+1)$ topological insulator model for \KD fermions. The bulk theory
consists of a topological gravity theory of Chern-Simons type \cite{Chamseddine:1990gk, Witten:2007kt, Zanelli:2005sa} whose boundary contains a massless four dimensional reduced \KD field gauged under a $U(1)$ symmetry which is inherited from the de Sitter gauge symmetry of the bulk gravity theory.

\section{\label{z4gauge} Gauging $Z_4$}

Quite generally the presence of
a 't Hooft anomaly can be understood as representing an obstruction to gauging
global symmetries. With this in mind it is instructive
to see how one would go about gauging the
residual $Z_4$ global symmetry discussed in the last section. Again, we will carry
out this procedure directly in the lattice theory. To do this we must first generalize
the boundary operator given in eqn.~\ref{boundaryop} in such a way the
action is invariant under a 
local $Z_4$ transformation of the \KD field corresponding to
\begin{equation}
    \phi(C_p)\to e^{i\frac{\pi}{2}\left(-1\right)^p n(C_p)}\phi(C_p)\quad n(C_p)=0,1,\ldots, 3
\end{equation}
This can be done if the incidence matrices $I(C_p,C_{p-1})$ introduced in
eqn.~\ref{incident} are generalized to $Z_4$ gauge fields $U(C_p,C_{p-1})$ that
transform under gauge transformations as 
\begin{equation}
 U(C_p,C_{p-1})\to e^{-i\frac{\pi}{2}\left(-1\right)^p n(C_p)}\,U(C_p,C_{p-1}) e^{i\frac{\pi}{2}\left(-1\right)^{p} n(C_{p-1})}
\end{equation}

In this way a locally $Z_4$ invariant massless
action can be constructed.~\footnote{It is not hard to take this prescription applied 
to a cubical cell decomposition of a regular hypercubic lattice, and show that the fields
$U(C_p,U_{p-1})$ become the usual gauge links of a staggered fermion.} 
However, to check for gauge invariance of the full
quantum theory we also need to examine the measure. For a single \KD field
there are two fermion integrations per $p$-simplex $\int d\phi(C_p)d\bar\phi(C_p)$ and 
in general this changes by an element of $Z_2$ under a local $Z_4$ transformation.
However, it should be clear that the measure can be made 
locally invariant if the system contains
multiples of two \KD fields. 

Since 't Hooft anomalies are RG invariants a non-zero U.V 
anomaly generically requires either massless composite fermions or Goldstone bosons to be present
in the spectrum of the low energy theory.~\footnote{Another
possibility is that the I.R phase corresponds to some sort of
non-trivial topological field theory.} This implies that
only theories with vanishing U.V 't Hooft anomalies can have a trivial
gapped state in the I.R that characterizes a massive symmetric phase.

\section{A non-perturbative mixed anomaly}

We have shown that
theories of \KD fermions in even dimensions possess a discrete
global $Z_4$ symmetry on coupling to gravity. Furthermore we inferred that the
system possesses a 't Hooft anomaly which arises if we try to gauge this $Z_4$
symmetry. Cancellation of this anomaly dictates that the theory must contain multiples of
two \KD fields.
We can now ask whether this theory possesses any additional anomalies.
In particular we are interested in possible mixed 't Hooft anomalies that arise
if we attempt to gauge any additional global symmetries. 

The staggered fermion theory in flat space is clearly invariant under a reflection symmetry
that inverts one lattice direction~.\footnote{Note that 
in Euclidean space such a reflection symmetry and time
reversal invariance are equivalent.}. 
In the continuum a 
reflection operation corresponds to changing from a right to a left handed coordinate system and hence reverses the orientation. 

We can now ask what happens if we now consider gauging the
reflection symmetry ? This means allowing for
the freedom to choose the orientation locally on the space. On an
orientable space it is possible to choose a single orientation for
the entire space and this corresponds to a constant (trivial) background for
the corresponding $Z_2$ orientation gauge field.
However, on a non-orientable space this is not possible. Propagation on such a space
then corresponds to choosing a non-trivial background for the associated
orientation gauge field. Notice that these statements
can be applied equally to a triangulation of the space.
One can ask whether the use of such a background breaks any global
symmetries of the theory. To answer this we will consider a 
triangulation of a particular non-orientable
space -- the
real projective plane $RP^D$. This has Euler
characteristic $\chi=1$ corresponding to the fact that on $RP^D$ the 
\KD operator possesses just a single
zero mode (a 0-form). In fact the real projective plane is obtained by identifying
antipodal points on the sphere $RP^D\sim S^D/Z_2$.
The question is what happens to the partition function of a discrete 
\KD fermion propagating on
such a triangulation ?

Let us decompose
the \KD field into two reduced fields $\Phi^1,\Phi^2$. 
To allow for possible $Z_4$ invariant four fermion terms
we couple these fields to a scalar field $\sigma$. To generate four fermion
terms one would need to add additional terms quadratic in $\sigma$ to the action 
but the argument
we give below holds robustly for {\it any} action which is even
in $\sigma$ including actions containing scalar kinetic terms. The fermion operator is given by
\begin{equation}
M=\delta^{ab}{\cal K}+\sigma(x)\epsilon^{ab}.\end{equation}
We will assume that the total action
is invariant under a discrete symmetry which extends the fermionic $Z_4$ discussed
in the previous section:
\begin{align}
    \Phi^a&\to i\Gamma \Phi^a\\
    \sigma&\to -\sigma.
\end{align}
Notice that this fermion operator is antisymmetric and real and hence all eigenvalues
of $M$ lie on the imaginary axis. The partition
function is then given by the Pfaffian ${\rm Pf}\,(M(\sigma))$ where we
will define the latter as the product of the eigenvalues of $M(\sigma)$
in the upper half plane in the background of some reference configuration
$\sigma=\sigma_0$. By continuity we define the Pfaffian to be the product of these same eigenvalues under fluctuations of $\sigma$ and, in particular, under the $Z_4$ transformation $\sigma\to -\sigma$. The question that then arises is whether the Pfaffian is invariant
under $Z_4$.
 
At first glance it seems all is well since it is easy
to prove that
\begin{equation}
\Gamma M\left(\sigma\right)\Gamma=-M\left(-\sigma\right)\label{pfaff}.\end{equation}
This result shows that the spectrum and hence the determinant is indeed invariant under 
the $Z_4$ transformation $\sigma\to -\sigma$. 
But this is not enough to show the Pfaffian itself is unchanged since 
there remains the possibility that an odd number of
eigenvalues flow through the origin as $\sigma_0$ is deformed smoothly to $-\sigma_0$ leading to a sign change. 
To understand what happens we consider a smooth interpolation of $\sigma$:
\begin{equation}
\sigma(s)=s\sigma_0\quad {\rm with\;}\quad s \in \left(-1,+1\right).\end{equation}\\
The question of eigenvalue flow can be decided by focusing on the behavior of
the eigenvalues of the fermion operator closest to the origin at small $s$.
In this region the eigenvalues of smallest magnitude
correspond to zero modes of the \KD operator. There is a single such mode
on $RP^D$ which satisfies the 
eigenvalue equation:
\begin{equation}
\sigma_0 s\epsilon^{ab}v^b=\mu v^a.\end{equation}
The two eigenvalues $\mu=\pm i\sigma_0 s$. Clearly these eigenvalues change sign 
as $s$ varies from positive to negative values which indeed leads to a Pfaffian sign change. 
This can also be seen
explicitly from eqn.~\ref{pfaff} since 
\begin{equation}
{\rm Pf}\,\left[M(-\sigma)\right]={\rm det}\,\left[\Gamma\right]{\rm Pf}\,\left[M(\sigma)\right]=-{\rm Pf}\,\left[M(\sigma)\right].\end{equation}
We thus learn that the Pfaffian corresponding to two reduced \KD fields or equivalently
one full \KD field
indeed changes sign under the $Z_4$ transformation. On integration
over $\sigma$ the 
value of any $Z_4$ invariant function of $\sigma$, including the partition
function itself, would then yield zero rendering expectation values of such operators
ill-defined. This corresponds to a non-perturbative mixed 
anomaly between the $Z_4$ and reflection symmetries. 
Our method of derivation
is similar to that used by Witten in discussing a non-perturbative
anomaly for odd numbers of Weyl fermions in $SU(2)$ \cite{Witten:1982fp}.

Again, we see that
this anomaly can be cancelled for systems
possessing multiples of two \KD fields since then eigenvalues flow through the origin
in pairs and the sign of the partition function does not change. This is true for a variety
of four fermion interactions since one can always perform orthogonal rotations on 
$2N$ reduced fermions to put the Yukawa interaction in the canonical form $(\lambda_1i\sigma_2\oplus\lambda_2i\sigma_2\oplus\ldots\oplus\lambda_Ni\sigma_2)$

For staggered fermions in flat space one should think of this
breaking of $Z_4$ as the manifestation of a mixed 't Hooft anomaly arising
as a result of gauging the reflection symmetry. 

We learn from these arguments 
that the minimal model of staggered fermions with no 't Hooft anomalies,
which is hence capable of symmetric mass generation,
contains two staggered or equivalently four {\it reduced} staggered fields.
In the continuum limit such a lattice theory gives rise to four or eight
Dirac fermions in two and four dimensions respectively which matches the number of
fermions which are needed to cancel 
off the discrete fermion parity and spin-$Z_4$ anomalies
of Weyl fermions in two and four dimensions respectively \cite{Garcia-Etxebarria:2018ajm}.

\section{Pati-Salam model on the lattice}
As an application of these results we will discuss a possible construction of
the Pati-Salam GUT model using staggered fermions. We start by considering a four
dimensional continuum
theory of four massless \KD fields in which the $Z_4$ 't Hooft anomaly discussed in the 
previous section vanishes.
The flat space action separates into two independent pieces $S=S_1+S_2$ 
where 
\begin{align}S_1&=\int d^4x\, 
{\rm Tr}\left[\Psib_+\gamma_\mu\partial_\mu\Psi_-\right]+\nonumber\\
S_2&=\int d^4x\,{\rm Tr}\left[\Psib_-\gamma_\mu\partial_\mu \Psi_+\right]\end{align}
where each term depends on four {\it reduced} \KD fields that transform under separate global $SU(4)$ symmetries and the trace 
in this expression
extends over both the $SU(4)$ and internal matrix indices for each reduced field.
Suppressing the $SU(4)$ indices and adopting a Euclidean chiral basis for the Dirac
matrices \begin{equation}
\gamma_\mu=\left(\begin{array}{cc}0&\sigma_\mu\\\overline{\sigma}_\mu&0\end{array}\right)\end{equation}
with $\sigma_\mu=(I,i\sigma_i)$) and $\overline{\sigma}_\mu=(I,-i\sigma_i)$
we find that the reduced fermion field $(\Psib_+,\Psi_-)$ appearing in $S_1$ can be written in the block form 
\begin{equation*}
\Psi_-=\left(\begin{array}{cc}
0&\psi_R\\
\psi_L&0\end{array}\right)\quad \Psib_+= \left(\begin{array}{cc}\psib_L&0\\0&\psib_R\end{array}\right)\end{equation*}
Evaluating the (internal) trace the action becomes:
\[S=\int d^4x\,\left[
{\rm tr}\,\left(\psib_R\overline{\sigma}^\mu \partial_\mu \psi_R\right)+
{\rm tr}\,\left(\psib_L\sigma^\mu \partial_\mu\psi_L\right)\right]
\]
We see that
each $2\times 2$ block corresponds to a doublet
of Weyl spinors 
transforming under a $SU(2)\times SU(2)$ flavor symmetry in addition to
the $SU(4)$ symmetry. 
Thus $\Psi_-$ contains the spinor 
representations $(4,2,1)_L\oplus (4,1,2)_R$ which makes
explicit the fact that the two blocks transform
under separate $SU(2)$ flavor groups.
In Minkowski space 
we can then trade the right handed fermions in the usual way for left handed fermions
via $\psi_R=i\sigma_2\psi^{\prime *}_L$. After doing this
it is clear that the reduced \KD field decomposes into a set of left-handed
Weyl fields in the representations
$(4,2,1)\oplus (\overline{4},1,2)$. Thus both the symmetries and representations of
this reduced \KD field  {\it in flat space} clearly match those
of the Pati-Salam GUT model \cite{Pati}.

However this connection to Pati-Salam is lost if we add in the other sector corresponding
to the reduced field $(\Psib_-,\Psi_+)$. To recover the Pati-Salam model we must ensure
that this {\it mirror} sector decouples from low energy physics by adding
suitable interactions that are capable of gapping just this sector {\it without} breaking
any symmetries either explicitly or spontaneously.
To avoid spontaneous symmetry breaking we require that all 't Hooft anomalies
vanish - which from our previous arguments will hold here because the
fermions come in multiples of four. To avoid explicitly breaking the symmetry
we add 
a $Z_4$ symmetric four fermion term of the form
\begin{align}
    \frac{G^2}{2}\int d^4x\,\epsilon_{abcd}\,
   \left[
    \tr (\Psib_-^a\Psib_-^b)  \tr (\Psib_-^c\Psib_-^d)+\right. \nonumber\\
    \left. \tr (\Psi_+^a\Psi_+^b)  \tr (\Psi_+^c\Psi_+^d)
   \right]
\end{align}
where we have made make explicit the $SU(4)$ indices in the mirror sector
and retain only a trace
$\tr$ over internal $SU(2)\times SU(2)$
indices. Equivalently we can introduce a scalar field $\phi$ 
that transforms in the real six dimensional
representation of $SU(4)$ and rewrite the four fermion term as a Yukawa interaction
\begin{equation}
    \delta S=G\int d^4x\,\hat{\phi}_{ab}\tr\left[\Psib_-^a\Psib_-^b+\Psi_+^a\Psi_+^b\right]
\end{equation}
where $\hat{\phi}$ satisfies
\begin{equation}
    \hat{\phi}_{ab}=\frac{1}{2}\left(\phi_{ab}+\frac{1}{2}\epsilon_{abcd}\phi_{cd}\right)
\end{equation}
Using this interaction we might hope to generate masses for the mirrors for large
coupling $G$ due to the formation of a condensate of mirror $SU(4)$ baryons.
One immediate objection to this strategy for gapping the mirror sector is that
any four fermion interaction corresponds to a perturbatively irrelevant operator and hence should not be capable of changing the I.R behavior
of a theory. However, numerical work on a related Higgs-Yukawa model with $SO(4)$ rather
$SU(4)$ symmetry
provides evidence that it this may be possible non-perturbatively with a symmetric
four fermion condensate forming for sufficiently strong coupling \cite{Butt:2018nkn,Butt:2021brl}. 

Furthermore, it may be possible to
evade the question of whether such four fermion 
terms are relevant by instead gauging the $SU(4)$ symmetry of the mirror sector. In that
case confinement can drive the mirror sector into a symmetric
gapped phase with four fermion condensate even for small Yukawa
coupling - a scenario that has been advocated by Tong et al. in the
context of chiral fermions \cite{Razamat:2020kyf}. Again,
there is evidence in favor of this
scenario from lattice simulations of a gauged $SO(4)$ Higgs-Yukawa
model \cite{Butt:2021koj}. 

It should be clear at this point how to construct a lattice mirror model that targets
the Pati-Salam theory in the naive continuum limit -- simply replace the continuum
\KD field $\Psi$ by a staggered lattice field $\chi$ transforming in the fundamental
representation of an $SU(4)\times SU(4)$ symmetry
together with suitable Yukawa interactions targeting a reduced
component of the full staggered field. In detail the proposed lattice action is
\begin{align}
    S&=\sum_{x,\mu}\eta_\mu(x)\left[\chib_+\Delta_\mu \chi_- +
    \chib_-\Delta^c_\mu\chi_+\right]+\\
    &G\sum_x \hat{\phi}_{ab}\left[\chib_-^a\chib_-^b+
    \chi_+^a\chi_+^b\right]+\frac{1}{2}\sum_x\hat{\phi}_{ab}^2
\end{align}
and the lattice covariant difference operator in the mirror sector is given by
\begin{equation}
    \Delta^c_\mu \chi_+(x)=U_\mu(x)\chi_+(x+\mu)-U^\dagger_\mu(x-\mu)\chi_+(x-\mu)
\end{equation}
with $U_\mu(x)$ an $SU(4)$ lattice gauge field. The mirror sector should confine
and produce a four fermion condensate for any non-zero $G$ without breaking
symmetries while the low energy sector should be a free theory whose continuum limit
corresponds to sixteen chiral fermions transforming in the Pati-Salam representations
of an $SU(4)\times SU(2)\times SU(2)$ global symmetry~\footnote{The current proposal
differs from that given in \cite{Catterall:2020fep} where the staggered field transforms
in the eight dimensional representation of a spin(7) group with a different set of
Yukawa interactions.}.  Notice that it is not possible to write down couplings of the
Pati-Salam and mirror sectors that are $SU(4)$ gauge invariant so that the Pati-Salam
sector should remain completely decoupled from the mirror sector in flat space.

\section{Conclusions}

In this paper we have shown that staggered fermions
experience both perturbative and non-perturbative gravitational
anomalies. To see these anomalies
we need to generalize staggered fields to discrete curved spaces by promoting
them to \KD fields. The perturbative anomaly breaks an exact global $U(1)$ symmetry
of the massless theory down to $Z_4$ in even dimensions. If we try to gauge this
$Z_4$ symmetry we detect the presence of a further 't Hooft anomaly. Cancellation of
this anomaly requires the theory to contain multiples of two \KD fields. This constraint
can also be seen in the presence of a mixed anomaly 
between the global $Z_4$ symmetry and a reflection symmetry when the latter is
gauged by considering the propagation of \KD fields on
non-orientable triangulations. It is quite remarkable that these anomalies
are manifest even in these discrete systems and place direct constraints on the continuum
infrared behavior of these theories. 

Similar exact lattice 
't Hooft anomalies have been
found for central branch Wilson fermions in two dimensions \cite{10.1093/ptep/ptaa003}. In this case it is again an exact staggered $U(1)$
symmetry, similar to the usual staggered $U_\epsilon(1)$ encountered
in this analysis that plays a crucial
role.

In the continuum limit two staggered fields
give rise to four or eight
Dirac fermions in two and four dimensions respectively which matches the number of
fermions which are needed to cancel 
off the discrete fermion parity and spin-$Z_4$ anomalies
of Weyl fermions in two and four dimensions respectively \cite{Garcia-Etxebarria:2018ajm}. 
In odd dimensions the story is similar. Staggered fermion theories with four
fermion interactions experience a 't Hooft anomaly associated with the $Z_4$
symmetry that can be cancelled only for multiples of two staggered fields.
In one and three dimensions such theories yield eight and sixteen
Majorana fermions in the continuum limit which again matches the number
needed to cancel 't Hooft anomalies 
arising from time reversal invariance. It appears that the constraints
arising from canceling gravitational
anomalies of \KD fermions match those associated with the vanishing
of a variety of discrete anomalies for Weyl or Majorana fermions.

As discussed earlier the cancellation of all 't Hooft anomalies is a necessary condition for
symmetric mass generation -- the appearance of a trivial gapped phase in theories
of interacting fermions and  the constraints described in this
paper agree with results for
gapping boundary states in topological superconductors 
\cite{You:2014vea, RevModPhys.88.035001}. A general review that summarises and synthesizes
these results can be found in \cite{Wang:2022ucy}. 
The cancellation of these anomalies also explains recent numerical studies
of symmetric mass generation with staggered
fermions \cite{Butt:2018nkn,Butt:2021brl}.

Finally we show how to construct
a simple continuum \KD theory satisfying these constraints that separates into a mirror
sector which can be decoupled from low energy physics and a light sector whose matter representations and global
symmetries match those of the Pati-Salam GUT model. It is interesting that considerations of gravitational
anomaly cancellation for \KD fermions yield a well known GUT model. Furthermore, the \KD construction suggests the existence of a massive mirror or dark sector whose
vacuum consists of a condensate of $SU(4)$ baryons together with
massive excitations that couple only gravitationally to the Pati-Salam fields in the low energy sector.
Clearly more work is needed to clarify whether these features can be exploited to construct
models of relevance to cosmology.

Finally the fact that the anomaly structure survives intact under discretization
naturally leads to a proposal
for a staggered
fermion model that targets the Pati-Salam GUT in the continuum limit.

\section*{Acknowledgments}
This work was supported by the US Department of Energy (DOE), 
Office of Science, Office of High Energy Physics, 
under Award Number DE-SC0009998. 

\bibliographystyle{unsrt}
\bibliography{refs}

\end{document}